\documentstyle[12pt]{article}

\def\ir{{\rm I}\hskip-.2em{\rm R}}
\def\half{\textstyle{\frac{1}{2}}}

\def\iZ{{\rm Z}\hskip-.5em{\rm Z}}
\def\ra{\rightarrow}
\def\tint{{\textstyle\int}}
\def\d{\partial}

\def\b{\begin{eqnarray*}}     
\def\e{\end{eqnarray*}}       
\def\bn{\begin{eqnarray}}     
\def\en{\end{eqnarray}}       
\def\<{\langle}
\def\>{\rangle}

\def\{{\lbrace}
\def\}{\rbrace}
\begin{document}
\title{Isolation and Expulsion of Divergences \\ in Quantum Field Theory}
\author{John R. Klauder\\
Departments of Physics and Mathematics\\
University of Florida\\
Gainesville, Fl  32611}
\maketitle
\begin{abstract}
Divergences that arise in the quantization of scalar quantum field models by
 means of a lattice-space functional integration may be attributed to a
single integration variable, and this fact is demonstrated by showing that
if the integrand for that single integration variable is appropriately
changed, then a perturbation expansion becomes order-by-order finite and
divergence free. The paper concludes with a brief review of a current
proposal of how an auxiliary, nonclassical potential added to the lattice
action of a relativistic scalar field quantization may automatically render
an analogous change of the integrand, and thus may lead, as well, to
nontrivial and divergence-free results.
\end{abstract}
\section{Introduction and Overview}
Traditional formulations of quantum field theory often encounter one or
another divergence in the course of calculation. For a self-interacting,
relativistic scalar field, for example, the divergences are often classified
on the basis of those that are encountered in the course of a perturbation
analysis. If we let $n$ denote the number of space-time dimensions, then a
typical classical action functional is given by
 $$I=\int\{\half[\d_\mu\phi(x)]^2-\half
m^2\phi(x)^2-g\phi(x)^p\,\}\,d^n\!x$$
appropriate to a $p^{\rm th}$ power interaction term. This is the action for
the so-called $\varphi^p_n$ model, and we shall confine attention to such
models. Here, there is a sum over the indices $\mu$ from $0$ to $s=n-1$, and
the metric signature is $[+1,-1,-1,\ldots,-1]$. The kind of divergences that
arise in quantization depend strongly on the space-time dimension $n$. For
example, for $n=2$, the only divergences that arise have to do with normal
ordering which amounts to a (re)definition of the local product for
operators. For $n\geq3$ the kind of divergences includes those covered by
normal ordering and generally additional divergences as well. In particular,
whenever $p<n/(n-2)$
it follows that normal ordering cures all divergences on a term by term
basis within a perturbation expansion. When
$p<2n/(n-2)$
we are in the so-called superrenormalizable regime, in which case infinite
renormalization counterterms are generally required, such as a divergent
mass renormalization counterterm for the $\varphi^4_3$ model. When
  $p=2n/(n-2)$
we are in the strictly renormalizable case where there is generally both
coupling constant and wave function renormalization. Finally, when
  $p>2n/(n-2)$
we are in the nonrenormalizable situation and the number of divergent
interaction counterterms that need to be added grows without limit. This
brief survey indicates the normal situation viewed from the point of view of
perturbation theory. For further details see \cite{boo}.

One may also formulate things differently. If we choose a Euclidean lattice
space functional integral formulation, then on the basis of the
renormalization group, for example, and outside of perturbation theory, one
can show that the nonrenormalizable cases, i.e., where $p>2n/(n-2)$ actually
pass to a {\it free} theory, or at least a generalized free theory, in the
continuum limit. Thus for the very important case where $p=4$, it follows
that for $n\geq5$ the theory becomes trivial in the continuum limit
\cite{fro}. Although the full proof has not yet been established there is
every indication that the same conclusion actually holds in the case $p=4$,
$n=4$. Thus we observe that different calculational schemes may very well
lead to distinct answers relative to the question of divergences.

Let us recall the appearance of the Euclidean-space lattice formulation for
the $\varphi^p_n$ model. We let
$k=(k_0,k_1,\ldots,k_s)\;,\;k_j\in\iZ\equiv\{0,\pm1,\pm2,\ldots\}$, denote a
lattice site, ${k^*}$ denote any one of the nearest neighbors (in the
positive direction) to the point $k$, and $a$ denote the lattice spacing.
Then the generating function for the Euclidean-space correlation functions
on the lattice is given by
  \b S(h)=C_N\int\exp\{\!\!\!\!&&\!\!\!\!\Sigma h_k\phi_ka^n-\half
Y(\phi_{k^*}-\phi_k)^2a^{n-2}\\ &&-\half
m_0^2\Sigma\phi_k^2a^n-g_0\Sigma\phi^p_ka^n\,\}\,\Pi\,d\phi_k\;.  \e
This integral is to run over $N\equiv(2L+1)^n$ lattice sites of a
(hyper)cubic lattice, and the continuum limit is one in which $a\ra0$ as
well as $L\ra\infty$ in such a way, ultimately, that the volume of the
space-time integration region becomes all of $\ir^n$. The constant $C_N$ is
fixed by the requirement that $S(0)=1$.
Here we have introduced unknown parameters which are to be determined, if
possible, so that a consistent and nontrivial (non-Gaussian) theory emerges
in the continuum limit. It is this formulation which, even allowing for
arbitrary choices of the parameters $Y>0,\,m_0^2$, and $g_0\geq0$,
nevertheless results in a trivial, Gaussian behavior in the continuum limit,
i.e., when $a\ra0$, for $n\geq5$ $(n=4)$. These results emerge from an
application of the renormalization group and/or the random-walk
representation \cite{fro}.

In the present paper we wish to sketch an argument which leads to a
qualitative appreciation of various divergences and which at the same time
{\it isolates a single parameter} which, in a manner of speaking, can be
said to be `responsible' for {\it all the divergences}. This identification
is significant because in work reported elsewhere \cite{kla}---and very
briefly summarized in the final section of the present paper---we have shown
how one may effectively eliminate an analogous parameter in lattice
formulations of more physical models. This procedure, when carried out to
its potential---a task, unfortunately, that has yet not been fully
accomplished---has the capability of offering a quantum field theory the
perturbation theory of which is {\it order-by-order divergence free!}
\section{Hyper-extreme Spherical Coordinates}
The lattice-space functional integral involves an integration over $N$
fields on the $N$ distinct lattice sites. Let us reexpress that integral in
{\it new coordinates} which we shall refer to as {\it hyper-extreme
spherical coordinates}. To that end, consider the measure
\b
\Pi\,d\phi_k\!\!\!\!&=&\!\!\!\!\Pi\,d\phi_k\,\delta(\kappa^2-\Sigma\phi_k^2)
\,d\kappa^2 =\delta(\kappa^2-\Sigma\phi_k^2)\,d\kappa^2\,\Pi\,d\phi_k\\
&=&\!\!\!\!\kappa^{N-2}\,d\kappa^2\,\delta(1-\Sigma\eta_k^2)\,\Pi\,d\eta_k
=2\kappa^{N-1}\,d\kappa\,\delta(1-\Sigma\eta_k^2)\,\Pi\,d\eta_k \;.\e
In the last line we have introduced $\phi_k\equiv\kappa\eta_k$ for all sites
$k$ in the lattice. Observe that $\kappa$ represents the {\it radius} of all
the $N$ fields while $\{\eta_k\}$ denotes the {\it direction field} of the
$N$ field amplitudes $\{\phi_k\}$. The direction field satisfies
$\Sigma\eta_k^2\equiv1$, and thus this field lies on (the surface of) the
$(N-1)$-sphere, $S^{N-1}$. Let us introduce these variables into the lattice
regularized functional integral. It follows (with $C_N$ absorbing the factor
$2$) that
  \b S(h)=C_N\int\exp\{\kappa\!\!\!\!\!\!&&\!\!\!\Sigma
h_k\eta_ka^n-\half\kappa^2Y\Sigma(\eta_{k^*}-\eta_k)^2a^{n-2}-\half
m_0^2\kappa^2\Sigma\eta_k^2a^n\\
&&-g_0\kappa^p\Sigma\eta_k^pa^n\,\}\,\kappa^{N-1}\,d\kappa\,\delta(1-\Sigma
\eta_k^2)\,\Pi\,d\eta_k\;. \e
Now observe that there are two qualitatively different kinds of integration
variables. On the one hand, there are the $\eta_k$ variables constrained so
that the sum of their squares lies on the unit sphere. Such variables are
quite `tame' in their behavior, and as we shall see they are not responsible
for the divergences even though there are a great many of them. On the other
hand, there is the single variable $\kappa$ which enters into the integrand
in one factor with a large power, namely the term $\kappa^{N-1}$. If we look
at the behavior of the $\kappa$ integral we see that it is largely
controlled by that term with a high power $(N-1)$. A steepest descent
integration for $\kappa$ is suggested and leads to the fact that
$\kappa\propto\sqrt{N}$ whenever $p$ and $n$ satisfy $p\leq2n/(n-2)$ (see
below). Not only does the steepest descent approach lead to a large value
for $\kappa$, it also leads to a relatively very narrow width of order one.
Of course, a change of variable given by rescaling $\kappa$ could put the
new support of order one, but then it would be extremely sharply
concentrated. {\it Essentially, it is the fact that the support of $\kappa$
is extremely sharply concentrated that ultimately leads to a divergent
perturbation expansion}. Let us illustrate and thereby clarify this remark.

Consider first the case of a free field in $n$ dimensions; specifically we
consider
  $$ S(h)=C_N\int\exp\{\kappa\Sigma
h_k\eta_ka^n-\half\kappa^2\Sigma(\eta_{k^*}-\eta_k)^2a^{n-2}-\half
m^2\kappa^2a^n
\}\,\kappa^{N-1}\,d\kappa\,d\sigma(\eta) $$
where we have set $Y=1$ and $m_0=m$, as is appropriate for this example.  We
have also explicitly used the fact that $\Sigma\eta_k^2=1$ in the
coefficient of the mass term, and introduced the abbreviation
  $$d\sigma(\eta)\equiv\delta(1-\Sigma\eta_k^2)\,\Pi\,d\eta_k\;.$$
We also introduce the notation for an average over $\sigma$ given by
$\<(\cdot)\>\equiv\int(\cdot) d\sigma(\eta)/\int d\sigma(\eta)$. First we
observe that $\<\eta_k^2\>\equiv(1/N)$ because every axis is treated
identically. For all $n\geq2$, we next observe that
$0\leq\<(\eta_{k^*}-\eta_k)^2\>\simeq(1/N)$ as well, as follows from the
previous equality plus the Schwarz inequality. The implication of this fact
is that there may be relatively little correlation between nearest neighbor
values. [On the other hand, for $n=1$, it follows that
$\<(\eta_{k^*}-\eta_k)^2\>\simeq (a/N)$, indicating some significant
correlation between nearest neighbor $\eta$ values.] We assume, and will
later confirm, that the gradient term is generally the dominant term in the
exponent. With that assumption it follows that the entire gradient term is
of order $N$, i.e.,
    $$\half\kappa^2\Sigma(\eta_{k^*}-\eta_k)^2a^{n-2}\simeq N\;.$$
This relation holds for the simple reason that each of the $N$ terms in that
sum will be of order one just as would be the integrand for any single
($N$-independent) integral. The value of $\kappa$ is well approximated by
the result of a steepest descent approximation, namely as given by
  $(N/\kappa)= \kappa a^{n-2}$
where we have already used the (upper limit of the) estimate for
$(\eta_{k^*}-\eta_k)^2$ and set $N-1\simeq N$. We conclude that
$\kappa\simeq\sqrt{Na^{2-n}}$. With this choice for $\kappa$ it follows that
the entire gradient term is $O(N)$ as desired. With these expressions we
also learn that
  $$\half m^2 \kappa^2a^n\simeq m^2 (Na^{2-n})a^n\simeq Na^2$$
which is of lower order than the gradient term justifying our neglect of it
in performing the steepest descent approximation. For comparison of relative
terms we may also examine a `unit volume' where $Na^n=1$. In that case, when
$n\geq3$, the gradient term is $O(a^{-n})$ while the mass term is
$O(a^{2-n})$; for $n=2$ the latter term should be interpreted as
 $a^0\equiv-\ln(ma)$ for some mass $m$. Thus for all $n\geq2$ the mass
 term is divergent as $N\ra\infty$ although it diverges at a slower rate
than the gradient term. Let us see what is the consequence of the divergence
 of the mass term.

Suppose we want to consider the relation of $S(h)$ for one mass term
compared to that for another by means of a perturbation in the difference
$\Delta m^2$ of the two masses. In particular, consider
   \b S(h)\!\!\!\!\!\!\!\!&&=C_N\int\exp\{\kappa\Sigma
h_k\eta_ka^n-\half\kappa^2\Sigma(\eta_{k^*}-\eta_k)^2a^{n-2}\\
&&\;\;\;\;\;\;\;\;\;\;\;\;\;\;\;\;\;\;\;\;\;\;-\half (m^2+\Delta
m^2)\kappa^2a^n
\}\,\kappa^{N-1}\,d\kappa\,d\sigma(\eta)\\
&&=C_N\int\exp\{\kappa\Sigma
h_k\eta_ka^n-\half\kappa^2\Sigma(\eta_{k^*}-\eta_k)^2a^{n-2}\\
&&\;\;\;\;\;\;\;\;\;\;\;\;\;\;\;\;\;\;\;\;\;\;-\half m^2\kappa^2a^n
\}(1-\half\Delta
m^2\kappa^2a^n+\cdots)\,\kappa^{N-1}\,d\kappa\,d\sigma(\eta)
\;.   \e
The point to observe in this calculation is that the term proportional to
$\Delta m^2$ is, as compared to the first term of the series, proportional
to $Na^2$ due to the presence of $\kappa^2a^n$ in the expansion. Thus we see
that a perturbation in the mass is automatically divergent as $N\ra\infty$
whenever $n\geq2$, a fact already well known in conventional treatments of
relativistic perturbation theory. Observe further that this divergence
arises from the {\it single} variable $\kappa$ and is not a consequence of
integrations over the many $\eta_k$ variables.

Let us next consider the case of a quartic perturbation term, namely
  \b S(h)\!\!\!\!&&\!\!\!\!=C_N\int\exp\{\kappa\Sigma
h_k\eta_ka^n-\half\kappa^2\Sigma(\eta_{k^*}-\eta_k)^2a^{n-2}\\
&&\;\;\;\;\;\;\;\;\;\;\;\;\;\;\;\;\;\;\;\;\;\;-\half
m_0^2\kappa^2a^n-g_0\kappa^4\Sigma\eta_k^4a^n
\}\,\kappa^{N-1}\,d\kappa\,d\sigma(\eta) \;. \e
Once again we assume to begin with that the gradient term is the most
dominant term in the exponent, an assumption that leads to the same estimate
for $\kappa$, namely $\kappa\simeq\sqrt{Na^{2-n}}$. The mass term is the
same as before, while the new term is the quartic coupling which becomes
  $$g_0\kappa^4\Sigma\eta_k^4a^n\simeq g_0(Na^{2-n})^2NN^{-2}a^n\simeq
Na^{4-n}\;.$$
In this expression the factor $1/N^2$ in the middle arises from the estimate
that $\<\eta_k^4\>\simeq 1/N^2$. In fact, this value can be determined from
  $$\<\eta_k^2\eta_l^2\>={1+2\delta_{kl}\over N(N+2)}$$
which holds exactly. Observe for $n<4$ that the quartic term is strictly
smaller than the gradient term, while for $n=4$ the two terms are of the
same order of magnitude. On the other hand, for $n>4$ the quartic term is
strictly larger than the gradient (or the mass term) and thus our assumption
of determining the behavior of $\kappa$ is unjustified. These three cases
correspond exactly to the three cases of superrenormalizable,
renormalizable, and nonrenormalizable fields in the usual perturbation
analysis \cite{boo}. Here we see how these three different situations appear
in the hyper-extreme spherical coordinates of the present paper.

As an example of estimation obtained in this picture, let us determine the
contribution to the vacuum polarization of the two-vertex bubble graph for
$n\geq3$ when estimated in hyper-extreme spherical coordinates. In
particular, let us estimate the expression
  $$C_N\int(\Sigma
:\kappa^4\eta_k^4:a^n)^2\exp\{-\half\kappa^2\Sigma(\eta_{k^*}-\eta_k)^2a^{n-
2}-\half m^2\kappa^2a^n\}\,\kappa^{N-1}\,d\kappa\,d\sigma(\eta)\;.$$ Recall
the usual prescription for normal ordering given by
  $$:\phi_k^4:\,=\phi_k^4-6\phi_k^2\<\phi_k^2\>+3\<\phi_k^2\>^2\;.$$
In terms of hyper-extreme spherical coordinates we see that
$$:\kappa^4\eta_k^4:\,=\kappa^4\eta_k^2-6\kappa^2\eta_k^2\<\kappa^2\eta_k^2
\>+3\<\kappa^2\eta^2_k\>^2\;.$$
Now we can use the fact that $\<\kappa^2\eta_k^2\>\simeq
Na^{2-n}N^{-1}=a^{2-n}$. All these terms then become of comparable value,
and to the accuracy we are working we cannot distinguish a cancellation of a
divergence (as with $\<:\phi_k^4\phi^4_l:\>=0$) from a noncancellation (as
with $\<:\phi_k^4::\phi^4_l:\>\not=0$). Rather we use the normal ordering to
establish the proper combinatorics of the correlations. Thus we determine,
at least to the correct order of magnitude, that the leading divergence of
the expression given above, relative to the unperturbed term, is given by
  $$(Na^{2-n})^4a^{2n}\Sigma\<\eta_k^4\eta_l^4\>\simeq
N^4a^{8-2n}a^{-n}NN^{-4}=Na^{8-3n}\;.$$
In the middle estimate, and apart from $\kappa^8$, we have used the fact
that the first sum would lead to a convergent integral were there a volume
element $(a^{-n})$, the second sum is translationally invariant $(N)$, and
the expectation of the $\eta$ variables involves $1/N^4$. The net result is
proportional to the volume $(N)$ as it should be. The remaining divergence
is ultraviolet as it should be. It is readily confirmed that the result is
appropriate for $n\geq3$.

The virtue of using hyper-extreme spherical coordinates is not necessarily
for calculational purposes, but rather to gain an overview of where, in some
sense, the divergences that appear in quantum field theory originate. With
the choice of hyper-extreme spherical coordinates we have exposed that the
radius $\kappa$---and especially its appearance in the integrand in the form
$\kappa^{N-1}$---is the source of all the divergences. Let us make this
point even more convincing by discussing what happens if that large power is
removed!
\section{Divergence-free Quantum Field Theory}
For the sake of discussion let us replace the term that we have identified
as the source of the divergences by something harmless. In particular, let
us replace $\kappa^{N-1}$ by $1$, i.e., effectively, in this term and {\it
in this term alone}, setting $N=1$. This is a very drastic change and it has
a number of consequences not the least of which is the loss of a local
interaction with which to make a relativistic model. We will comment on this
very point later, where we will briefly recall the principal argument, which
is presented elsewhere \cite{kla}, in which an analogous replacement appears
in such a way that locality in the continuum limit is retained. For the
present---and this is the main point of the present paper---we wish to
emphasize the profound consequences that arise when such a change is made in
a lattice space functional integral.

With the indicated change in the integration measure, the lattice space
generating functional of interest becomes
  \b S'(h)=C'_N\int\exp\{\kappa\!\!\!\!\!\!&&\!\!\!\Sigma
h_k\eta_ka^n-\half\kappa^2Y\Sigma(\eta_{k^*}-\eta_k)^2a^{n-2}-\half
m_0^2\kappa^2\Sigma\eta_k^2a^n\\
&&-g_0\kappa^p\Sigma\eta_k^pa^n\,\}\,d\kappa\,\delta(1-\Sigma\eta_k^2)\,\Pi
\,d\eta_k\;. \e
One immediately sees that the integration over the single variable $\kappa$
is normal and without any special feature. On the other hand, the parameters
of the model, i.e.,  $Y$, $m_0$, and $g_0$, need not behave as they
previously did, and indeed, the cutoff dependence of these parameters is one
of the interesting features that needs to be determined. To that end let us
first examine the special case in which $Y=0=g_0$ and only $m_0$ remains. In
that case
   $$ S'(h)=C'_N\int\exp\{\kappa\Sigma h_k\eta_ka^n-\half m_0^2\kappa^2a^n
\,\}\,d\kappa\,\delta(1-\Sigma\eta_k^2)\,\Pi\,d\eta_k\;.  $$
Although $\kappa>0$, the symmetry of the measure under $\eta\ra-\eta$ lets
us treat the integral over $\kappa$ as a simple Gaussian integral, and the
result (changing $C'_N$ as needed) becomes
  $$S'(h)=C'_N\int\exp\{\,\half m_0^{-2}(\Sigma
h_k\eta_k)^2a^n\,\}\,\delta(1-\Sigma\eta_k^2)\,\Pi\,d\eta_k\;.  $$
 The remaining integrals over the $\eta$ variables may be approximated by
simply replacing the factor $\eta_k\eta_l$ by its average value of
$\delta_{kl}/N$. When this is done the result becomes
 $$ S'(h)=\exp\{\,\half m_0^{-2}N^{-1}\Sigma h_k^2a^n\,\}\;,  $$
and in the continuum limit the only way for the result to be nontrivial is
that $m_0^2=m^2/N$, where $m^2$ is a cutoff independent quantity. This
dependence of the bare mass is of course quite different than in the usual
case. Furthermore, it is noteworthy that a perturbation in the mass term
leads to a series that is order-by-order {\it finite}. In particular,
consider the expression
  \b S'(h)\!\!\!\!\!\!\!\!&&=C'_N\int\exp\{\kappa\Sigma
h_k\eta_ka^n-\half(m^2+\Delta m^2)N^{-1}\kappa^2a^n
\,\}\,d\kappa\,\delta(1-\Sigma\eta_k^2)\,\Pi\,d\eta_k\\
    &&=C'_N\int\exp\{\kappa\Sigma h_k\eta_ka^n-\half m^2N^{-1}\kappa^2a^n
\,\}\\
&&\;\;\;\;\;\;\;\;\;\;\;\;\;\;\;\;\;\;\;\;\;\;\;\times(1-\half\Delta
m^2N^{-1}\kappa^2a^n+\cdots)\,d\kappa\,\delta(1-\Sigma\eta_k^2)\,\Pi\,d\eta_
k\;.  \e
The essential fact to observe about this equation is that the term of order
$\Delta m^2$ is actually $O(1)$ due to the fact that $\kappa^2$ is large of
order $Na^{-n}$. Thus there is no divergence in the perturbation series as
long as we choose the mass renormalization as indicated.

Next let us introduce a nonlinear coupling term. This leads to the
expression
  \b S'(h)=C'_N\int\exp\{\kappa\!\!\!\!\!\!&&\!\!\!\Sigma h_k\eta_ka^n-\half
m_0^2\kappa^2\Sigma\eta_k^2a^n\\
&&-g_0\kappa^p\Sigma\eta_k^pa^n\,\}\,d\kappa\,\delta(1-\Sigma\eta_k^2)\,\Pi
\,d\eta_k\;. \e
In view of the behavior of the integral over $\kappa$, it follows that this
expression will be well defined and contributing if and only if
  $$g_0=gN^{-1}a^{(p/2-1)n}$$
in which case the integral is equivalent (using $\kappa=\varrho
N^{1/2}a^{-n/2}$) to
  \b S'(h)\!\!\!\!\!\!\!\!\!&&=C'_N\int\exp\{\varrho N^{1/2}\Sigma
h_k\eta_ka^{n/2}-\half m^2\varrho^2\\
&&\;\;\;\;\;\;\;\;\;\;\;\;\;\;\;\;\;\;\;-g \varrho^pN^{p/2-1}\Sigma\eta_k^p
\,\}\,d\varrho\,\delta(1-\Sigma\eta_k^2)\,\Pi\,d\eta_k\;,  \e
where, as usual, the constant $C'_N$ has been readjusted to account for the
change of variable. A perturbation expansion in the nonlinear coupling
constant leads to
\b S'(h)\!\!\!\!\!\!\!\!\!&&=C'_N\int\exp\{\varrho N^{1/2}\Sigma
h_k\eta_ka^{n/2}-\half m^2\varrho^2\,\}\\
&&\;\;\;\;\;\;\;\;\;\;\;\;\;\;\;\;\;\;\times(1-g
\varrho^pN^{p/2-1}\Sigma\eta_k^p+\cdots)
\,d\varrho\,\delta(1-\Sigma\eta_k^2)\,\Pi\,d\eta_k\;,  \e
It follows, relative to the first term, that the first-order correction is
$O(1)$.
 A similar analysis shows that each of the terms in the perturbation series
is finite. In other words, the perturbation series is {\it order-by-order
finite}. It is particularly noteworthy in this regard that $g_0$ is {\it
small} (specifically $g_0\propto N^{-1}a^{(p/2-1)n}\,$), meaning that the
factors that it multiplies need to be allowed to become large in order for
the nonlinear term to contribute. A large value for the factors multiplying
the coupling constant is analogous to what happens in the usual case where
the factor $\kappa^{N-1}$ is present in the integrand. But not only are
these large, they also have a broad range of values, and that situation is
quite unlike what happens in the usual case when the factor $\kappa^{N-1}$
is present.

Finally we turn our attention to the gradient term. For this purpose we
first let $m_0=0=g_0$ so that we may focus on the essentials. The expression
of interest reads
 $$S'(h)=C'_N\int\exp\{\kappa\Sigma
h_k\eta_ka^n-\half\kappa^2Y\Sigma(\eta_{k^*}-\eta_k)^2a^{n-2}\,\}
\,d\kappa\,\delta(1-\Sigma\eta_k^2)\,\Pi\,d\eta_k\,. $$
Clearly the role of the gradient term is to introduce correlations into the
direction field that have been, in the absence of that term, lacking.
Nevertheless, since $\eta_k^2\simeq 1/N$, we observe, as previously noted,
that $0\leq(\eta_{k^*}-\eta_k)^2\simeq1/N$ as a consequence of the Schwarz
inequality. Thus to make this term a contributor it is necessary that
  $$\kappa^2Y\Sigma(\eta_{k^*}-\eta_k)^2a^{n-2}\simeq 1$$
a condition that requires that
  $$ Y\simeq (Na^{-n})^{-1}a^{2-n}\<\Sigma(\eta_{k^*}-\eta_k)^2\>^{-1}\simeq
N^{-1}a^2\;.$$

Finally, in this regard, we combine the several terms needed to make up the
full theory (minus, of course, the factor $\kappa^{N-1}$ in the integrand!).
Based on the fact that each of these terms has a bounded contribution, it
follows that we may choose as our basic lattice-space expression
  \b S'(h)=C'_N\int\exp\{\kappa\!\!\!\!\!\!&&\!\!\!\Sigma
h_k\eta_ka^n-\half\kappa^2N^{-1}a^2\Sigma(\eta_{k^*}-\eta_k)^2a^{n-2}-\half
m^2N^{-1}\kappa^2a^n\\
&&-gN^{-1}a^{(p/2-1)n}\kappa^p\Sigma\eta_k^pa^n\,\}\,d\kappa\,\delta(1-
\Sigma\eta_k^2)\,\Pi\,d\eta_k\;. \e
This equation, with the indicated cutoff-dependent coefficients, represents
our proposal for obtaining a nontrivial (and generally non-Gaussian)
continuum limit in the special case that the factor $\kappa^{N-1}$ is
removed from the integrand.

Moreover---and this is of central significance---{\it the form of the
renormalization of coefficients involved is entirely multiplicative and in
no way involves normal ordering}. This has the general consequence that for
a covariant quantum field theory defined in some way which is analogous to
the present discussion, {\it subtractive} renormalizations would {\it not}
be required; instead, all required renormalizations would be {\it strictly
multiplicative}.

Let us now turn our attention to a very brief discussion of how we believe a
scenario that is schematically similar to that of the present paper may well
be adapted to a relativistic self-interacting scalar field.
\section{Application to Relativistic Models}
The quantization of any system entails ambiguities including the
introduction of additional potential terms to the Hamiltonian operator or to
the action that are proportional to $\hbar$. In field quantization it is
generally acknowledged that some such additions are necessary. The proposal
we have in mind adds an auxiliary, $\hbar$-dependent potential $P$ to the
lattice action designed to dramatically alter the continuum limit.
Specifically, the potential $P$ is designed to make the ground-state
distribution of the field at a sharp time have a generalized Poisson
distribution in the continuum limit. One way to arrange this is for the
distribution to be chosen already on the lattice to have a generalized
Poisson distribution. To illustrate this proposal we choose the
characteristic function for the lattice-space, sharp-time field distribution
$|\Psi(\phi)|^2$ to have the form given by
   $$\int e^{i\Sigma'
g_k\phi_ka^s}|\Psi(\phi)|^2\,\Pi'\,d\phi_k=\exp\{-\tint[1-e^{i\Sigma'
g_k\phi_ka^s}]\rho(\phi)\,\Pi'\,d\phi_k\}  $$
where $\rho>0$. Here, $g_k$ denotes a test sequence at a constant lattice
time, and the sums and products run over the set of lattice sites at a
constant lattice time.
To ensure that the model has a unique translationally invariant state it is
necessary and sufficient that $\int\rho(\phi)\,\Pi'\,d\phi_k=\infty$, which
implies that we deal with a generalized Poisson distribution \cite{def}.
This divergence can be traced to a necessary singularity in the integrand
for small field values which is accommodated by the ansatz that
$$\rho(\phi)={J^2\,e^{-2W(\phi)}\over\Pi'_k[\Sigma'_l\,\beta_{k-l}\phi_l^2]^
\gamma}\;.$$
Here $J^2$ is a normalization constant, $W(\phi)$, with $W(0)=0$, accounts
for the intermediate and large field behavior and leads to the finiteness of
the moments, $\{\beta_k\}$, $k$ a lattice point, is a set of positive
constants (e.g., $\beta_k\propto\exp[-\Sigma_{j=1}^s|k_j|]\,$) subject to
the normalization condition $\Sigma'_k\beta_k=1$ where the sum and product
are over all $N'=(2L+1)^{s}$ lattice sites at a constant lattice time and
with periodic boundary conditions, and finally
$\gamma=1/2+\xi/(2N'),\;0<\xi<1$. With this implicit ansatz for the lattice
ground state $\Psi(\phi)$, the lattice Hamiltonian is defined by
  \b {\cal H}\!\!\!\!\!\!\!\!&&=-{1\over2}Y^{-1}a^{-s}\sum\biggl[{\d^2\over
\d\phi_k^2}-
{1\over\Psi(\phi)}{\d^2\Psi(\phi)\over\d\phi_k^2}\biggr]\\
   &&\equiv -{1\over2}Y^{-1}a^{-s}\sum{\d^2\over\d\phi_k^2}+{\cal
V}(\phi)\;.  \e
Although it is generally difficult to determine an analytic expression for
the ground state $\Psi(\phi)$, it is possible to argue that in the continuum
limit the local form of the auxiliary potential $P$ is proportional to
$\phi(x)^{-2}$ with a coefficient proportional to $\hbar^2$. By using the
freedom to choose $W$ we hope to ensure that the remainder of the potential
$\cal V$ has the necessary lattice form for a covariant continuum limit. Of
course, once the proper Hamiltonian is in hand, it is possible, in
principle, to build various correlation functions.

We finally comment on the relation between the proposal of the present
section for a relativistic field and the discussion in the foregoing
sections. That connection is readily made when we once again introduce
hyper-extreme spherical coordinates, this time on a {\it spatial slice} of
the lattice. In particular, the form for the characteristic functional for
the sharp-time field becomes
$$\exp\{-2J^2\!\tint[1-e^{i\kappa\Sigma'g_k\eta_ka^s}]e^{-2W(\kappa\eta)}\,
\Pi'_k[\Sigma'_l\beta_{k-l}\eta_l^2]^{-\gamma}\!\kappa^{-1-\xi}d\kappa\,
\delta(1-\Sigma'\eta_k^2)\Pi'd\eta_k\}\,.  $$
Observe that the factor $\kappa^{-1-\xi}$ is a result of a quotient between
a factor $\kappa^{N'-1}$ in the numerator which arises from the measure and
a factor $\kappa^{2\gamma N'}=\kappa^{N'+\xi}$ that appears in the
denominator. The cancellation of the large power $\kappa^{N'}$ that arises
is exactly analogous to the dropping of the factor $\kappa^{N-1}$ in the
previous section of this paper! Lastly, we note that the choice of the
resultant power, i.e., $-1-\xi$, is designed to give a divergent result for
$\int\rho(\phi)\Pi'd\phi_k$, and yet, with the help of $W$, make every
moment of the smeared, sharp-time field $\Sigma'
g_k\phi_ka^s=\kappa\Sigma'g_k\eta_ka^s$ finite.
\section*{Dedication}
It is a pleasure as well as an honor to dedicate the present article to the
memory of Hiroomi Umezawa, whose warmth and enthusiasm in all matters,
scientific and otherwise, was universally recognized and appreciated.

\end{document}